# ChatISA: A Prompt-Engineered, In-House Multi-Modal Generative AI Chatbot for Information Systems Education

**Fadel M. Megahed**

Information Systems and Analytics
Miami University
0000-0003-2194-5110

**Ying-Ju Chen**

Mathematics
University of Dayton
0000-0002-6444-6859

**Joshua A. Ferris**

Information Systems and Analytics
Miami University
0009-0007-3202-1331

**Cameron Resatar**

Information Systems and Analytics
Miami University
0009-0005-5502-2084

**Katie Ross**

Information Systems and Analytics
Miami University
0009-0008-8999-6276

**Younghwa Lee**

Information Systems and Analytics
Miami University
0000-0002-1322-2214

**L. Allison Jones-Farmer**

Information Systems and Analytics
Miami University
0000-0002-1529-1133

## Abstract:

As generative AI ('GenAI') continues to evolve, educators face the challenge of preparing students for a future where AI-assisted work is integral to professional success. This paper introduces ChatISA, an in-house, multi-model AI chatbot designed to support students and faculty in an Information Systems and Analytics (ISA) department. ChatISA comprises four primary modules: Coding Companion, Project Coach, Exam Ally, and Interview Mentor, each tailored to enhance different aspects of the educational experience. Through iterative development, student feedback, and leveraging open-source frameworks, we created a robust tool that addresses coding inquiries, project management, exam preparation, and interview readiness. The implementation of ChatISA provided valuable insights and highlighted key challenges. Our findings demonstrate the benefits of ChatISA for ISA education while underscoring the need for adaptive pedagogy and proactive engagement with AI tools to fully harness their educational potential. To support broader adoption and innovation, all code for ChatISA is made publicly available on GitHub, enabling other institutions to customize and integrate similar AI-driven educational tools within their curricula.

# 1 Introduction

"Despite the potential benefits of pedagogical (generative AI) chatbots, IS empirical studies of pedagogical chatbot efficacy in higher education are limited and fewer still discuss their potential challenges, drawbacks, impacts on equity and accessibility, or threats to privacy, justice, and fairness" (Chen et al., 2023, p. 162).

As the potential of generative AI continues to evolve, educators face an urgent need to equip students for a future in which AI-assisted work is not only prevalent but may also become essential to professional success. Generative AI, deep learning models capable of generating high-quality text, images, code, data analytics outputs, and other forms of content based on their training datasets (IBM Research, 2023), has catalyzed an unprecedented wave of adoption across both industry and higher education. From the introduction of the Transformer architecture (Vázquez-Cano, 2021) to the launch of ChatGPT on November 30, 2022 (OpenAI, 2022), and the development of multimodal models such as GPT-4o, Claude 3.7 Sonnet, and Gemini 2.5 Pro (Under, 2025), generative AI capabilities have expanded at an exponential rate in recent years. These technologies have demonstrated remarkable proficiency in natural language understanding, reasoning, and text generation, driving widespread adoption across domains including customer service, software development, and education. This rapid proliferation is reflected in the global generative AI market, which was valued at $67 billion in 2024 and is projected to increase dramatically to $967 billion by 2032 (Fortune Business Insights, 2025).

In response to the pressing need to understand the values and impact of generative AI, the Information Systems (IS) community has recently advanced scholarly discourse on the topic. Notably, the Journal of the Association for Information Systems (JAIS) published a special issue on "Generative AI and Knowledge Work" (Alavi et al., 2024), while several flagship IS journals have issued calls for papers addressing various aspects of generative AI. These include *MIS Quarterly*'s call for research on "Artificial Intelligence–Information Assurance Nexus: Shaping the Future of Information Systems Security, Privacy, and Quality" (Chen et al., 2024) and *Decision Support Systems*' forthcoming issue on "Generative AI: Transforming Human, Business, and Organizational Decision Making" (Zhang et al., 2024). Of relevance to education, *Communications of the Association for Information Systems* has called for contributions to its special issue on "Digital Transformation of Higher Education: Pedagogical Innovation Toward Education 4.0" (Abbas et al., 2024). This initiative aligns with the broader vision of Education 4.0, a paradigm shaped by Industry 4.0 technologies, including artificial intelligence, to facilitate personalized, experience-based learning.

Within the context of generative AI's impact on education, scholars have emphasized the importance of equipping students with meta-learning skills (Maudsley, 1979), which enable them to adapt to continuous change, acquire new competencies, and effectively apply their knowledge across diverse contexts (Bisgambiglia et al., 2024). Given that the mission of Information Systems in business education is to prepare students to effectively leverage technology and data for organizational success, IS departments should play a pivotal role in equipping students for a professional landscape in which generative AI-assisted tasks are increasingly prevalent.

Several initiatives have sought to integrate generative AI into higher education (e.g., Chen et al., 2023; Chukwuere & Handoko, 2024). For instance, Chukwuere and Handoko (2024) highlighted the transformative impact of generative AI on student learning experiences while identifying key challenges such as academic integrity concerns, limitations in understanding user inputs, and issues related to resource allocation. While these studies offer valuable insights into the educational applications of generative AI, several critical gaps warrant further investigation.

First, prior research has predominantly focused on general-purpose generative AI models rather than in-house generative AI systems designed with domain-specific large language models (LLMs) and proprietary datasets. General-purpose LLMs suffer from fundamental limitations that affect their reliability, adaptability, and real-world applicability. Since these models are trained on vast amounts of internet-based text, they inherently reflect historical biases, inaccuracies, and context-independent knowledge, which can lead to hallucinations and erroneous responses (Lin et al., 2024). These challenges primarily arise from their lack of grounding, a critical process that connects LLM-generated outputs to external knowledge, contextual constraints, and real-world dynamics. The absence of explicit mechanisms for integrating real-time, domain-specific knowledge limits their suitability for high-stakes decision-making contexts, such as legal reasoning, financial analytics, and medical diagnosis (Echterhoff et al., 2024).

In response to these concerns, enterprises have accelerated the development of in-house generative AI applications, which leverage proprietary datasets and domain-specific knowledge bases to enhance accuracy. These systems often employ Retrieval-Augmented Generation (RAG), fine-tuning, or reinforcement learning to improve outputs. Notable examples include Clara, KPMG's Audit AI; Omnia, Deloitte's Financial AI; and Leah, PwC's legal service. These models are designed to address industry-specific challenges and improve trustworthiness, underscoring the potential value of similar approaches in academic settings.

Second, despite its widespread use in sectors such as e-commerce, finance, healthcare, media, manufacturing, and supply chain management, much of the existing literature on generative AI in education has focused on non-business disciplines, particularly in non-IS contexts. Only recently have studies begun exploring the application of generative AI in business education. For example, Riapina (2024) proposed a conceptual framework for generative AI-enabled business communication, while Yang and Evans (2019) developed AI-driven conversational chatbot prototypes for business simulation games. Additionally, Chen and Anyanwu (2025) examined the effectiveness of Moodle-powered AI chatbots in Nigerian business education, finding that they significantly enhanced student learning. However, further research is needed to explore in-house generative AI tools designed using rigorous design science methodologies to effectively support business students' academic and professional development.

Third, most studies on generative AI in education have focused on a single LLM model, despite the increasing availability and use of multiple LLMs by students and educators (Kelly, 2024). Different LLMs exhibit varying strengths and weaknesses, with no single model consistently excelling across all tasks. Consequently, students and educators often interchangeably use multiple LLMs to compare outputs, identify optimal solutions, or supplement learning. Given this complementarity among LLMs, developing generative AI systems that integrate and leverage the strengths of multiple models is essential for enhancing educational effectiveness.

Finally, while developing in-house generative AI systems and integrating them into IS curricula is a critical priority for IS scholars, such a system has yet to be introduced within the IS community. The system is essential for advancing education and mitigating disparities in access to advanced AI technologies, often called the "AI divide." While some generative AI tools are freely available, enabling broader access, many state-of-the-art (SOTA) models remain restricted due to high costs.

For example, as of February 24, 2025, the extended thinking mode of the Claude 3.7 Sonnet model is unavailable in its free-tier chat interfaces (Anthropic, 2025). Additionally, priority access, new feature rollouts, and higher rate limits (e.g., 5x higher in Claude's case) are typically reserved for paid Plus and Pro plans offered by OpenAI and Anthropic (Claude, 2024). This AI divide mirrors the well-documented digital divide, the gap between individuals with and without access to modern digital technologies. Studies by organizations such as the Organisation for Economic Co-operation and Development (OECD) (Gottschalk & Weise, 2023) and the Public Policy Institute of California (Starr et al., 2022) have highlighted how such disparities limit educational opportunities, particularly for underprivileged groups. If left unaddressed, the AI divide could exacerbate inequities in higher education, hindering students from fully leveraging AI-assisted learning. Given the rapid adoption of in-house generative AI systems in industry, the 124% increase in AI-related job postings from 2023 to 2024 by ZipRecruiter (Kaufman, 2025), and the dramatic rise in student use of generative AI tools (e.g., 90% of Harvard students utilize generative AI) (Hirabayashi et al., 2024), IS educators must take proactive steps to integrate AI technologies into curricula.

To address these challenges and offer IS students hands-on experience with cutting-edge AI technology, we designed ChatISA—an in-house, multi-model, modular AI chatbot. ChatISA is trained using multiple LLMs, proprietary datasets, and prompt engineering to assist students in IS education. This paper documents ChatISA's design, implementation, and impact, provides open-source Python code for replicability, presents examples of its use in two Information Systems and Analytics (ISA) courses, and discusses broader implications for generative AI in IS education.

## 2   Literature Review

### 2.1   Generative AI Chatbot

During the early 2020s, the generative AI chatbot industry witnessed significant growth with the emergence of platforms such as ChatGPT and Bard, both of which leverage the Transformer Neural Network Architecture. This architecture enhances natural language processing by efficiently managing complex and lengthy input sequences through selective attention mechanisms, enabling the models to learn intricate language patterns and relationships and generate contextually relevant and coherent responses (Al-Amin et al., 2024). A generative AI chatbot is an advanced conversational system powered by natural language processing (NLP) and deep learning, allowing it to produce text that closely resembles human writing in both style and coherence (Nguyen et al., 2022). These chatbots are capable of generating real-time, human-like responses in conversations and exhibit cognitive functions such as problem-solving, contextualized reasoning, prediction, planning, and decision-making (Kim et al., 2025). As a result, generative AI chatbots have been widely implemented across various business sectors to support tasks such as responding to frequently asked questions, providing troubleshooting assistance, facilitating marketing efforts, and managing service inquiries (Chen et al., 2023).

In the field of education, generative AI chatbots have emerged as innovative educational tools, enabling them to serve as partners, assistants, and mentors throughout students' learning processes. Researchers have increasingly focused on exploring their potential impact across various academic disciplines, including medicine, psychology, and language studies (Jeon, 2024). Several studies have highlighted the positive contributions of AI chatbots to education. For example, Chukwuere and Handoko (2024) emphasized the transformative role of generative AI chatbots in higher education, noting their ability to enhance student learning, support research activities, and streamline administrative tasks. Similarly, Ilieva et al. (2023) suggested that AI chatbots can foster critical thinking by assisting students with academic tasks such as idea generation, research, and writing. Their study further emphasized that interactive conversations with AI chatbots encourage students to critically analyze information and engage with diverse perspectives. Grassini (2023) underscored the significance of the round-the-clock availability of generative AI chatbots, which ensures continuous academic support for students and faculty. This accessibility benefits distance learners and individuals with varied schedules, learning preferences, and academic expectations. Furthermore, Wu and Yu (2024) found that AI chatbots increase student motivation, engagement, and self-esteem while reducing anxiety and stress. By offering 24/7 academic assistance, these tools ultimately enhance students' overall academic performance.

Despite these benefits, prior research has also identified several challenges associated with the adoption and use of generative AI chatbots in higher education. Key concerns include academic integrity issues, as AI-generated responses may raise plagiarism and authenticity concerns (Ilieva et al., 2023). Privacy and security risks are also prevalent, as AI chatbots may inadvertently store or misuse sensitive student data (George & Wooden, 2023). Additionally, AI-generated responses may suffer from hallucination, leading to factually incorrect or misleading information (Chukwuere & Handoko, 2024). Resource allocation constraints further complicate the widespread adoption of these technologies, particularly for institutions with limited infrastructure or funding. Moreover, enforcing uniform institutional policies and regulations regarding generative AI usage remains challenging, as universities struggle to develop standardized guidelines that ensure ethical and responsible use of these tools (Neumann et al., 2023). Ethical concerns also persist, particularly regarding bias in AI-generated content and the potential over-reliance on automated educational support (Chen et al., 2023).

To optimize the benefits of generative AI chatbots in education while mitigating associated challenges, recent research has focused on developing in-house chatbot solutions tailored to institutional needs. Reddy et al. (2024) proposed a framework for designing an AI chatbot specifically for educational institutions to facilitate student access to academic information and administrative services. However, their study did not extend to implementing a functional system. In another study, Rienties et al. (2024) introduced an institutional AI digital assistant, i-AIDA, and compared its effectiveness to general-purpose AI assistants. Their findings suggest that i-AIDA provides more effective student support while ensuring academic integrity, improving accuracy, and enhancing trust and security. Furthermore, the study highlighted i-AIDA's ability to protect intellectual property and user privacy. However, its findings have limitations, as students evaluated the system based on a five-minute recorded video rather than engaging in direct, hands-on interaction with the chatbot.

In summary, continued advancements in generative AI solutions are essential for improving students' academic experiences and preparing a highly skilled workforce for the AI-driven era. As these technologies become increasingly integrated into educational settings, IS scholars must play a pivotal role in their development and refinement, ensuring that AI-driven educational tools are designed to enhance learning outcomes while addressing ethical, security, and institutional concerns.

## 2.2    Task-Technology Fit Theory

The foundational theory underpinning the design and development of ChatISA is the Task-Technology Fit Theory (Goodhue & Thompson, 1995). This theory postulates that optimal use and performance benefits of technology are realized when the features of the technology align well with the task requirements. Figure 1 illustrates the principal constructs of this theory: task characteristics, technological characteristics, and task-technology fit. Task characteristics encompass the nature of tasks, such as complexity, structure, and interdependence, which define an individual's actions within specific settings (Gupta et al., 2024). Technology characteristics refer to the features and functionalities of computer systems, including hardware, software, and data, highlighting aspects like usability, compatibility, reliability, and relevance to the tasks at hand. Task-Technology Fit is defined as the degree to which technology supports individuals in performing their tasks, with a higher degree of fit predicting greater technology usage and improved task performance (Goodhue & Thompson, 1995, p. 216).

Over the past three decades, this theory has been employed to understand the impact of task-technology fit on the use of various technologies across different tasks. These include mixed reality-based training (Albeedan et al., 2023), digital technology adoption (Roth et al., 2023), and big data analytics (Wang & Lin, 2019). Empirical evidence supports the notion that when designed to effectively support task completion, technology can significantly enhance task performance and future use. Following the theoretical constructs of the Task-Technology Fit Theory, ChatISA has been meticulously designed to ensure that its functionalities effectively and efficiently support ISA students in their tasks. This alignment enhances task performance and promotes increased utilization of the system.

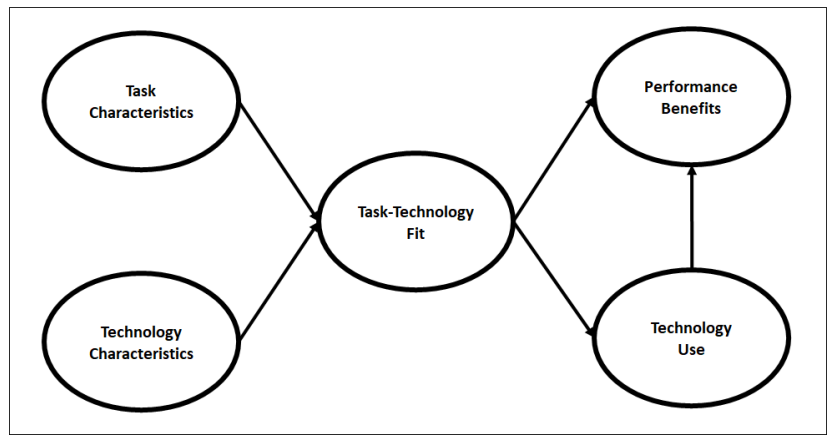

**Figure 1. General Model of Task-Technology Fit**

ChatISA (available at https://chatisa.fsb.miamioh.edu/) comprises four primary modules, each tailored to address different educational needs. The first module, *Coding Companion*, has been designed to assist students with coding-related inquiries, considering both their academic backgrounds and the preferred coding styles used within the ISA program. This program integrates multiple programming languages into its curriculum, including Python, R, and SQL. Additionally, students can create a collaborative coding experience by using the output from one model as input to another in an iterative manner. The second module, *Project Coach*, steers students through ISA project-related challenges. The third module, *Exam Ally*, has been designed to assist with exam preparation by generating questions based on information extracted from uploaded PDFs of ISA course materials (e.g., lecture notes or slides). The user specifies the type of generated questions through a dropdown menu. The fourth module, *Interview Mentor*, has been designed to improve technical interview readiness by generating tailored interview questions based on a user-input job description and an uploaded resume in PDF format. Across all four modules, students can choose from various LLMs offered by OpenAI, Anthropic, Mistral, and Cohere. We expect that the specific technology characteristics of ChatISA provide a better alignment for these tasks than general LLMs and other generative AI tools, resulting in enhanced task performance, increased satisfaction, and greater utilization.

# 3    ChatISA System

To develop ChatISA, we adopted a design science research approach as shown in Figure 2, leveraging the structured framework outlined by (Peffers et al., 2007). This methodology has been widely recognized for its effectiveness in developing innovative artifacts that address complex challenges (e.g., (Anderson et al., 2023)). The framework consists of six key stages: The framework involves six phases: *problem identification*, *defining solution objectives*, *artifact design and development*, *artifact demonstration*, *artifact evaluation*, and *communication*.

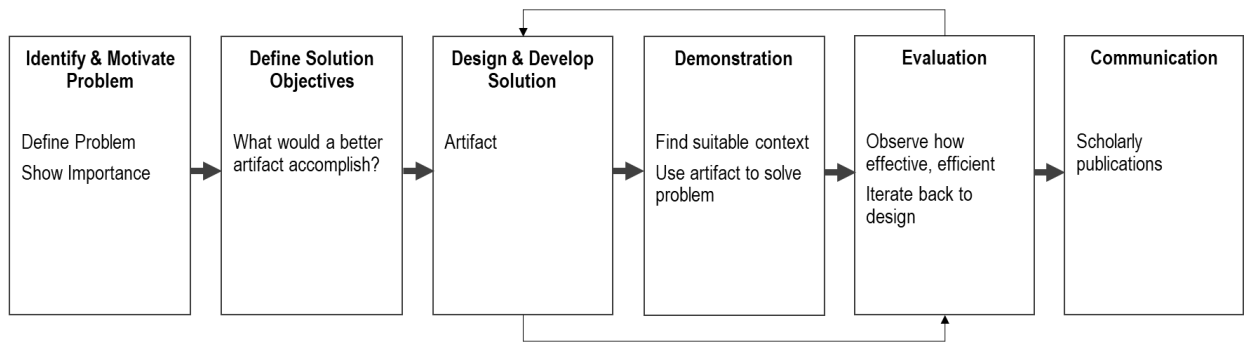

**Figure 2. The Design Science Research Methodology Applied to The ChatISA Development**

### 3.1 Problem Motivation of ChatISA

The initial stage of our design focuses on identifying the core problem: the slow and limited adoption of generative AI in business school classrooms. This delay restricts students' exposure to technologies increasingly vital for workforce readiness. As previously noted, the primary issue driving the development of ChatISA is the limited and slow adoption of generative AI tools within business school classrooms. This lag significantly restricts business students' opportunities to engage with technologies that have already been widely adopted across industry sectors and are increasingly critical for workforce readiness. A recent report by *Inside Higher Ed* indicates that only half of higher education institutions currently provide access to GenAI tools (Flaherty, 2025).

Furthermore, findings from a recent survey conducted by the Association to Advance Collegiate Schools of Business (AACSB) involving business school deans (n = 236) and faculty (n = 429) underscore this gap: while 55% of faculty reported using GenAI as a teaching or delivery tool and 63% integrated it into curriculum content, only 33% employed it for learning assessments, and a mere 28% used it to support students' hands-on learning activities and assignments (AACSB, 2025). In contrast, student demand is notably higher - 70% of recent graduates expressed a desire for GenAI integration into coursework, and more than half reported feeling unprepared for the workplace due to insufficient exposure (Coffey, 2024).

Although some schools, like the University of Toronto's business school, have pioneered tools like the "All Day TA", such efforts often depend on costly, large-scale LLMs, which present significant cost barriers (Murray, 2025). There is an urgent need for affordable, in-house generative AI (GenAI) tools, to ensure broader access and promote digital equity in business education, especially at resource-constrained institutions.

### 3.2 Define Solution Objectives of ChatISA

The next phase of our design process focuses on establishing solution objectives. The design requirements and principles for this study are informed by both a thorough analysis of the identified problem, insights gained from an extensive literature review, and inputs from ISA students and faculty. In formulating our design principles, we considered both user needs (ISA students) and organizational priorities (ISA department). As a result, our requirements emphasize key factors such as the fit between tasks and ChatISA functionality, usability, privacy, flexibility, and affordability.

#### 3.2.1 Fit

Prior research on technology adoption (e.g., Huy et al., 2024; Roth et al., 2023) suggests that users are more inclined to value technology when it supports the accomplishment of tasks they find personally meaningful or engaging. For example, Roth et al. (2023) found that the fit between the tasks of Germany's Federal Office and the capabilities of blockchain technology—namely transparency, reliability, adaptability, and security—significantly influenced the successful adoption of the blockchain system. Huy et al. (2024), through an online survey of 671 ChatGPT users, empirically demonstrated the positive impact of task-technology fit on its actual usage. In the context of our study, users' perceptions of task–technology fit, specifically, the degree to which AI chatbots' features and capabilities align with their primary tasks, may significantly influence their willingness to adopt the technology. Building on this insight, we engaged in iterative discussions with prospective users (i.e., ISA students) to identify the types of academic tasks they find most relevant and frequently performed. This process led to identifying four focal use cases detailed in Section 2.2: *Coding Companion*, *Project Coach*, *Exam Ally*, and *Interview Mentor*. We anticipate that ISA students will exhibit greater propensity to adopt the AI chatbot when they perceive a strong alignment between ChatISA's functionalities and their academic tasks, particularly if the tool demonstrably improves task performance in terms of efficiency and effectiveness.

### 3.2.2    Usability

According to ISO (International Organization for Standardization, 2018), usability is defined as "the extent to which a product can be used by specified users to achieve specified goals with effectiveness, efficiency and satisfaction in a specified context of use." This concept plays a critical role in driving widespread technology adoption and sustained user engagement. Recent scholarly literature and practitioner analyses suggest that AI systems with high usability are more likely to be integrated into users' everyday practices. For instance, based on testing results with ChatGPT, Nielsen (2023) emphasized the significant impact of usability, particularly learnability, on business professionals' productivity and the quality of their work. Similarly, Yeon et al. (2024), drawing on a survey of 500 users of generative AI services, identified trustworthiness as a critical usability-related interface feature influencing user perceptions and adoption intentions. Chen et al. (2021) also provided empirical evidence demonstrating that usability significantly affects both perceived value and user satisfaction in interactions with AI-based chatbots. Recognizing the importance of usability in shaping technology acceptance, we aimed to design ChatISA's user interface with key usability principles in mind. Specifically, we incorporated features that promote ease of learning, navigation, and interaction; a consistent visual design (e.g., font, typeface, and color scheme across pages); enhanced readability; user support functionalities (e.g., the ability to export chats to PDF); and trust-building elements (e.g., disclaimers, citations, developer information, and version history).

### 3.2.3    Privacy

As concerns over data privacy and security have emerged as significant barriers to the adoption of AI tools, regulatory frameworks such as the General Data Protection Regulation (GDPR) and the Health Insurance Portability and Accountability Act (HIPAA) have imposed stringent compliance requirements on the handling of sensitive data and system development practices. In response, organizations are increasingly developing in-house generative AI solutions or adopting "semi-open" on-premises deployment models that allow proprietary data to remain securely hosted on internal servers (Huang et al., 2025). For instance, following an incident involving the inadvertent leakage of confidential data to ChatGPT, Samsung prohibited the use of the tool and implemented stricter security protocols to govern its application (Park, 2023). Private AI infrastructures offer organizations greater control over their data, ensuring that sensitive information does not pass through public networks, thereby mitigating privacy risks and facilitating compliance with data protection mandates. In line with this approach, we designed our AI chatbot as an internally managed system, capable of securely storing and processing confidential academic materials, including assignments, exams, instructional content, team project data, and interview question repositories, as well as student profiles and interaction histories.

### 3.2.4    Flexibility

A growing body of industry and academic evidence highlights a shift toward the use of multiple task-specific large language models (LLMs) instead of relying solely on a single, general-purpose AI system. The *Financial Times* (Waters, 2025) reports that organizations increasingly adopt multi-LLM strategies, selecting models based on task fit, domain needs, and factors like cost, speed, and output quality. This approach also mitigates the risk of vendor lock-in by promoting flexibility and interoperability (Seifi & Chugh, 2025). On an individual level, users increasingly engage with multiple AI platforms, comparing outputs across models to identify the most accurate or context-appropriate response, thereby reducing risks to confidently presented yet inaccurate information (Kirrane, 2025). For example, Perplexity AI offers a user-friendly interface that aggregates responses from various LLMs, including GPT-4, Claude, and DALL-E 3, while also providing source information to enhance transparency (Garn, 2025). Within educational contexts, students have likewise demonstrated a preference for combining distinct AI tools based on functionality - employing one model for programming support, another for writing assistance, and yet another for research purposes (Kelly, 2024). Such behavior highlights the appeal of integrated multi-LLM platforms that consolidate diverse capabilities into a single system, reducing the need to toggle between separate tools.

Building on this trend, ChatISA incorporates multiple LLMs, including Claude, GPT, Gemma, LLaMA, and Command R+, to support varied academic tasks. It offers both large and lightweight variants (e.g., GPT-4o vs. GPT-4o-mini) so users can balance performance and efficiency. The interface allows seamless switching between models, enabling iterative exploration and enhancing both satisfaction and effectiveness.

### 3.2.5 Affordability

A recent *Inside Higher Ed* report highlights that one of the primary barriers to widespread adoption of generative AI in higher education is the substantial cost associated with the required hardware, software, energy consumption, and ongoing system maintenance (Flaherty, 2025). These resource demands contribute to widening digital inequities, particularly between well-funded private or flagship institutions and smaller, resource-constrained colleges and universities. At the individual level, access to premium generative AI tools such as ChatGPT Plus, which requires a $20 monthly subscription, remains limited for many students due to affordability constraints, despite the potential learning benefits offered by these advanced features. In response to these challenges, educators and institutions have taken proactive steps to ensure equitable access to AI tools by minimizing financial barriers and providing students with the necessary technological infrastructure to support learning. In alignment with these efforts, we designed and implemented ChatISA within a constrained departmental budget, aiming to maintain access to advanced AI capabilities at a manageable cost. From September 2023 to May 2025, the monthly operational cost of the system remained between $100 and $250, funded by the ISA department to support inclusive student access to generative AI technologies.

### 3.3 ChatISA Design, Development and Demonstration

The subsequent phase in the design science research methodology involves the design and demonstration of the proposed IT artifact. We developed ChatISA using Python (Python 3.12.9). While we used multiple Python libraries in the development, two libraries form the backbone of our application. We used *Streamlit* (1.42.2) for the web-based graphical user interface and to harness the power of its multipage app framework, which facilitates the integration of multiple modules, currently four within our application. We also used the open-source *LangChain* (0.3.19) library to have a consistent API for invoking and retrieving data from the APIs of multiple LLMs. It is important to note that since the APIs (as opposed to the Chat platforms) are stateless (i.e., they do not keep track of the previous messages by design), we manually append the previous messages (i.e., prompts and responses) to each user prompt to maintain the expected conversational experience in ChatISA. Figure 3 provides an overview of ChatISA's design as of version 2.0.0.

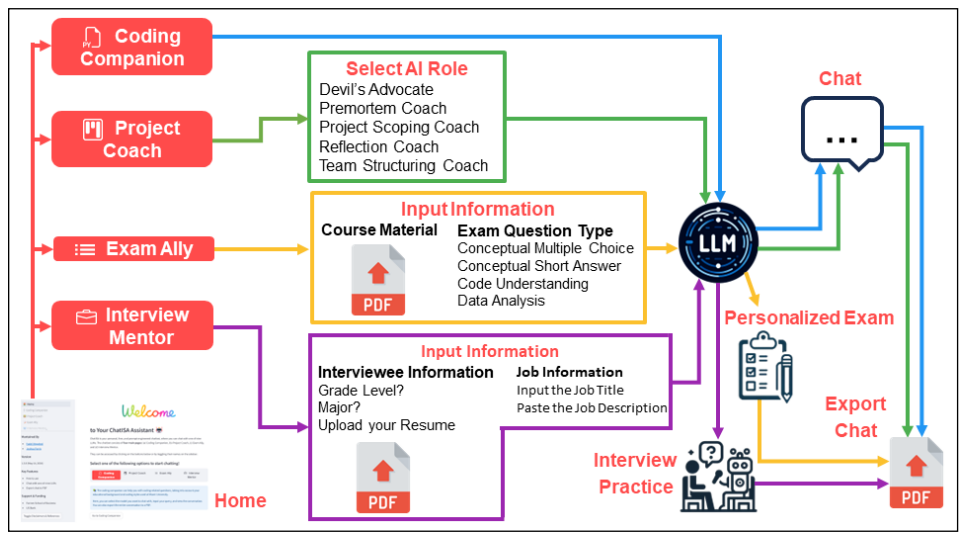

**Figure 3. An Overview of ChatISA's design as of Version 2.0.0**

As shown in Figure 4, each of the four modules has a similar look to simplify the students' experience. Specifically, on the left, we provide a sidebar, which presents information about the app, its current benefits and limitations, and a dropdown menu for LLM selection. To the right of the sidebar, the main window serves as the interaction area where users input their prompts, and the AI generates the corresponding message.

This design echoes OpenAI's approach, with the prompt input text bar at the bottom and the ongoing chat conversation above it.

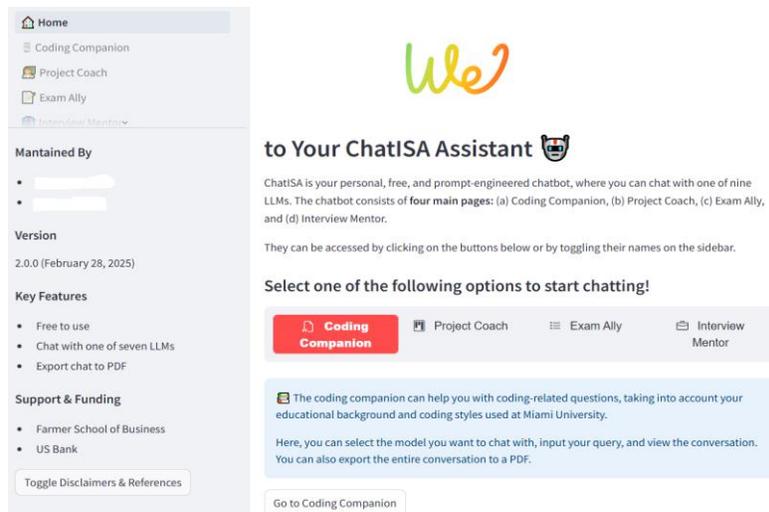

**Figure 4. ChatISA User Interface**

To improve user experience, we incorporated a custom PDF export feature within an interactive HTML container utilizing *Streamlit's* expander. When the container is expanded for the first time, students are prompted to enter their name and course number, which are incorporated into a neatly formatted PDF. The PDF includes a title page with an auto-generated title ("Student's Name's Interaction with ChatISA on Date"), followed by sections on "ChatISA's Purpose" (explaining its function and listing all LLMs used), "ChatISA's PDF Output Style and Layout" (describing how prompts, code, and text outputs are formatted), and "Token Counts and Cost Breakdown" (detailing input/output tokens and associated costs). The full chat transcript follows, concluding with our custom instructions. This approach to PDF generation is consistent across all four modules, keeping in mind that instructors may require students to document their interactions with LLMs for reference or to assess the degree of customization in student prompts to generate acceptable answers. For code and more details on our export to PDF functionality, we refer the reader to our GitHub repository (https://github.com/fmegahed/chatisa) in the supplementary materials (see "chatpdf.py" in the "lib" folder).

### 3.3.1 The Coding Companion

The *Coding Companion* module is designed to assist undergraduate students majoring in ISA with their coding-related inquiries. This module employs a system prompt to provide an interactive and supportive learning experience. The Coding Companion starts by introducing itself as the student's ChatISA Assistant, ready to help with any questions. It encourages students to specify the subject and topic they want to learn, ensuring that the assistance is tailored to their understanding and knowledge. Listing 1 shows the system prompt used in this module.

---

You are an upbeat, encouraging tutor who helps undergraduate students majoring in business analytics understand concepts by explaining ideas and asking them questions. Start by introducing yourself to the student as their ChatISA Assistant who is happy to help them with any questions.

Only ask one question at a time. Ask them about the subject title and topic they want to learn about. Wait for their response. Given this information, help students understand the topic by providing explanations, examples, and analogies. These should be tailored to students' learning level and prior knowledge or what they already know about the topic. When appropriate also provide them with code in both **R** (use tidyverse styling) and **Python** (use pandas whenever possible), showing them how to implement whatever concept they are asking about.

When you show **R** code, you must use:

(a) library_name::function_name() syntax as this avoids conflicts in function names and makes it clear to the student where the function is imported from when there are multiple packages loaded. Based on this, do NOT use **library**() in the beginning of your code chunk and use **if**(**require**(**library**)==**FALSE**) **install.packages**(**library**), and

(b) use the native tidy **|>** as your pipe operator.

On the other hand for **Python**, break chained methods into multiple lines using parentheses; for example, do NOT write **df.groupby**('Region')['Sales'].**agg**('sum') on one line.

---

**Listing 1. System Prompt for the Coding Companion Module in ChatISA**

To ensure consistency and adherence to our instructions, we set the temperature to 0 for all the LLMs that students can choose from the dropdown menu. These include: 'gpt-4o', 'gpt-4o-mini', 'claude-3-7-sonnet-20250219', 'command-r-plus', 'gemma2-9b-it', 'llama3.1-8b-INSTANT', and 'llama3.3-70b-versatile'. Setting the temperature to 0 (i.e., the least creative value) reduces the variability in the output and increases the likelihood that the LLM will follow our system prompt.

The primary pedagogical goal of the Coding Companion is to foster an interactive and personalized learning environment. Unlike existing LLMs that provide generic responses, the Coding Companion is designed to engage students actively by asking questions and tailoring explanations based on their responses. This approach ensures that the help aligns with the student's current understanding and learning needs. By breaking down complex concepts into manageable explanations and using examples and analogies, the Coding Companion aims to enhance comprehension and retention. Furthermore, the Coding Companion emphasizes the importance of practical implementation by providing code snippets in both R and Python. This dual-language approach acknowledges the diverse coding preferences within the ISA program and ensures students gain proficiency in both languages. The module's focus on using specific syntax and style guidelines for R (e.g., library_name::function_name() and the native pipe operator |>) and Python (e.g., breaking chained methods into multiple lines) helps students write clear, maintainable code, reinforcing some of our programming practices.

### 3.3.2   Project Coach

The *Project Coach* module is designed to guide students through various stages of their ISA team projects by assuming one of the five roles: Scoping Coach, Premortem Coach, Team Structuring Coach, Devil's Advocate, and Reflection Coach. Each role has a specific system prompt to ensure the interaction is tailored to the specific role. The system prompt for the Scoping Coach is detailed in **Error! Reference source not found.**. The prompts for the remaining four roles of the Project Coach were adapted from (Mollick & Mollick, 2023). Exact prompts are available in the link provided in the Supplementary Materials. Similar to the *Code Companion* module, we used a temperature setting of 0 and provided the students with the same set of models.

---

You are an AI assistant designed to interactively guide users through defining an analytics project using a project scoping document template. Your goal is to help the user provide detailed information for each section of the document, offer feedback, and refine their inputs to create a comprehensive project scope. Each section corresponds to a question in arabic numerals in the scoping document.

Here is the project scoping document template you will be working with:

{project_scoping_document}

To begin, ask the user to provide a short description of their project. Respond with the following response:

Thank you for choosing to scope your analytics project with me. To get started, please provide a brief description of your project in a few sentences.

Once the user provides their project description, respond with:

Great! Let's now walk through each section of the project scoping document to define your project in more detail. We'll start with Section 1 and work our way through the document. For each section, I'll ask you to provide the necessary information and offer feedback to help refine your inputs.

Then, iterate through each section of the project scoping document:

1. Identify the current section number and title.

2. Prompt the user to answer the required question for each section, providing hints and subquestions, **if** they were provided in the project scoping template. The user does not have the project scoping document so you need to provide them with each question.

3. Once the user provides their input, offer feedback and suggestions to help refine their response per the following guidelines:

   – reflect on the user's input and identify areas that may need clarification or improvement.

   – provide specific suggestions or questions to help the user enhance their input for the section.

4. Iterate with the user until the section is completed to mutual satisfaction.

5. Store the finalized answer for the section, along with its section number and information.

6. Move on to the next section and repeat steps 1-5 until all sections are completed.

For the timeline section, **if** the user's response is not very specific:

---

```
<timeline_guidance>

    – Help them identify the tasks and how they should be broken down over time.
    – Provide suggestions on creating a more detailed timeline.
  </timeline_guidance>

Once all sections of the project scoping document have been completed, present the user with the fully scoped project
    document in the following format:

Ensure that the final output mimics the structure of the uploaded project_scoping_document, including a mix of questions
    and answers. Provide an appropriate title on top of the final output, and summarize the answers in tables for the
    sections that include a table. Furthermore, use these guidelines to structure the final output:

<project_scope_output>
Congratulations! We have completed the project scoping document for your analytics project.
    Here is the final version:

Project Scope for [Insert Project Title]
<section1_question_and_answer>
<section2_question_and_answer>
<section3_question_and_answer>
<section4_question_and_answer>
<section5_question_and_answer>
<section6_question_and_answer>
<section7_question_and_answer>
<section8_question_and_answer>
<section9_question_and_answer>
<section10_question_and_answer>
</project_scope_output>

Thank you for taking the time to define your project in detail. This comprehensive project scope will serve as a valuable guide
throughout the execution of your analytics project. If you have any further questions or need assistance, please don't hesitate to
ask.
```

**Listing 2. System Prompt for the Project Scoping Coach[1]**

The primary pedagogical goal of the Project Coach is to foster a structured and interactive learning environment that enhances students' project management skills throughout the entire project lifecycle. The different roles correspond to various project stages, from initial scoping to final reflection. The Project Scoping Coach leverages a template created by one of the ISA faculty for her ISA class, guiding students through each step and helping them produce a comprehensive project scoping document. The other four roles: Premortem Coach, Team Structuring Coach, Devil's Advocate, and Reflection Coach are based on existing pedagogical research by Mollick and Mollick (2023). Premortem Coach enables project premortem exercises by inspiring teams to anticipate potential risks and their solutions; Team Structuring Coach promotes the identification and effective use of team resources and skills; Devil's Advocate questions ideas and assumptions at different project stages; and Reflection Coach aids teams in structured reflection, allowing them to identify lessons learned and insights from their project experiences. We contribute to integrating these prompts into a single tool that allows students to experiment with different AI roles using various LLMs. By providing a familiar, browser-based interface within the ChatISA app, we eliminate the need for students to set up and adapt to different platforms, thereby enhancing their learning experience and making it easier to leverage these tools throughout their projects.

### 3.3.3   Exam Ally

The Exam Ally module is designed to assist students in preparing for their exams by generating practice exam questions based on their uploaded course documents. These documents may include textbooks, lecture notes, or study guides. The system prompt ensures that the questions generated are aligned with the specific content and question style chosen by the student. Listing 3 provides the detailed system prompt. Here, we limited the models to the best available (i.e., frontier) models only: 'gpt-4-turbo' (which we have now replaced with 'gpt-4o'), 'claude-3-7-sonnet-20250219', 'command-r-plus', and the open-sourced 'llama3.3-70b-versatile.

---

[1] We utilized Anthropic's Prompt Generator tool (Anthropic, Prompt generator, 2024) to fine-tune our original prompt.

```
You will be acting as an AI tutor to help students prepare for an information systems and analytics exam. You will be provided
    with a course document that may be a textbook, lecture notes, or study guide. Your goal is to generate practice exam
    questions for the student based on: (a) their uploaded course document, and (b) their chosen exam question style.

#Course Document:
{course_text}

#Exam Question Style:
{exam_type}

    Generate exam questions using the following criteria:
        –   If the student requested short-answer questions, generate 10 questions that cover a range of topics from the
            documents.
        –   For all other question types, generate 4-5 questions.
        –   Show one question at a time and wait for the student to provide an answer before moving on to the next
            question.
    Once you receive the student's answer, acknowledge it and present the next question.
        Continue this process until the student has answered all of the questions you generated.

    After the student has completed all the questions, provide them with the following:

1.  In a <feedback> block, write a detailed evaluation of their performance on each question. Structure your feedback as
    follows:
        –   Begin with an overall summary paragraph highlighting areas of strength and areas for improvement.
        –   Then, go through each of their answers, first restating the question, then assessing the correctness and
            completeness of their answer. Provide guidance on how they could improve their answer if applicable.
        –   End with some motivating words of encouragement.
2.  After the <feedback> block, provide their overall exam score in a <score> block. Calculate the score as
    follows:
        –   Divide 100 points evenly across all questions (e.g. if there were 10 questions, each one is worth 10 points).
        –   Award points for each question based on the correctness and completeness of the student's answer.
        –   Sum up the points and provide a final score out of 100.

    Some key things to remember:
        –   Be patient and encouraging in your tone. The goal is to help the student learn and feel more prepared for their
            exam.
        –   Provide detailed and constructive feedback, pointing out both strengths and weaknesses in their answers.
        –   Generate questions that test a range of concepts and skills from the course documents.
        –   Do not show the student the questions in advance; they should be seeing them for the first time when you
            present them in a one-question-at-a-time format.
```

**Listing 3. System Prompt for the Exam Ally Module in ChatISA**

Unlike the previous two modules, we have set the temperature parameter to 0.25 to allow for some creativity in the exam questions while maintaining a balance with instructional guidance. Using a dropdown menu, the student can select one of the following exam/question types: "Conceptual Multiple Choice", "Conceptual Short Answer", "Code Understanding", or "Data Analysis". Additionally, this module requires the student to input a readable PDF of the relevant course materials before starting the chat conversation. We utilize the Python libraries *PyMuPDF* (1.25.3) and *pdf4llm* (0.0.9) to translate the input text to markdown, which is well-suited for LLM prompting (Artifex Software Inc., 2024). This translation process ensures that the LLMs accurately represent and interpret the content easily. These components are incorporated as part of constructing the f-string for the system prompt in Python, while the selected LLM is accounted for via *LangChain* and our chat generation function as in the previous two modules.

The primary pedagogical goal of the *Exam Ally* module is to help students assess their understanding of course material by generating practice questions based on their uploaded documents. This allows students to generate questions for every lesson or part of a lesson, facilitating continuous self-assessment. By presenting questions one at a time and providing detailed feedback, the Exam Ally helps students identify areas of strength and weakness, offering constructive guidance to improve their understanding. Additionally, the module leverages study guides and other course materials to create a comprehensive exam preparation experience, particularly useful for courses that do not utilize traditional textbooks. This approach aligns with the principles of affordable learning initiatives, e.g., *Ohio's Affordable Learning Initiative* (OhioLink, 2022), which promotes the use of free and low-cost educational resources. By supporting courses that rely on diverse types of materials, Exam Ally allows students to access effective study tools regardless of the resources used in their course. Furthermore, this tool can empower students to take ownership of their learning journey. If they feel that a lesson or part of a lesson was not well understood, they can use the chatbot to quiz their understanding of the topic.

### 3.3.4 Interview Mentor

The *Interview Mentor* module is designed to help students prepare for job interviews by generating tailored interview questions based on their resumes and the job descriptions they provide. The module leverages students' backgrounds to create relevant and meaningful interview questions. **Error! Reference source n ot found.** shows the system prompt for the *Interview Mentor*.

Similar to the previous module, we limited the models to frontier models only: 'gpt-4o', 'claude-3-7-sonnet-20250219', 'command-r-plus', and 'llama3.3-70b-versatile'. We also set the temperature to 0.25 to allow for some creativity while pushing the LLM to follow our system prompts. This module included the largest number of user inputs, with dropdown menus for model choice, grade and major, a PDF upload for the resume (translated to markdown using *PyMuPDF* and *pdf4llm*), and text inputs for job title and job description (which the student is asked to copy and paste from the job advertisement for the job they are applying for). Except for the LLM model choice, these inputs are passed to the system prompt via an f-string in Python. As before, the selected LLM is accounted for via *LangChain* and our chat generation function.

In addition to text-based interactions, the Interview Mentor also includes a voice recognition feature, allowing students to respond to interview questions verbally. This functionality supports more realistic interview practice, helps students become comfortable speaking their responses aloud, and simulates a live interview setting more closely.

The primary pedagogical objective of the *Interview Mentor* is to help students be more prepared for their interviews by providing a realistic and structured interview experience. Unlike online tools such as *Google's Interview Warmup* (Google, 2024), the *Interview Mentor* tailors the interview questions based on the student's resume, job ad, and background. With this customization, we hope that the questions generated are relevant and meaningful, addressing the specific requirements of the job and the student's qualifications. By structuring the order and type of questions, drawing from our experience and best practices from online tools like Google's Interview Warmup, the module attempts to provide comprehensive, effective, and customized interviews.

> You are an expert technical interviewer. You are interviewing a {grade} student, who is majoring in {major} for a {job_title} position.
>
> The student has provided you with their resume:
> {resume_text}
>
> The job description for the {job_title} position is as follows:
> {job_description}
>
> Carefully read and analyze the student's resume to understand their background and qualifications, and how it relates to the job description. Extract relevant information from the job description pertaining to the job duties and responsibilities, required qualifications, preferred qualifications, and any other relevant information.
>
> Once you have analyzed the resume and job description, conduct a structured interview with the student to assess their qualifications for the position. The interview should consist of six questions, asked one at a time.
>
> 1. Ask a background question about the student's interest in the position. Assess whether the candidate has a good understanding of the role, and has the necessary skills/drive to learn on the job.
>
> 2. Ask a background question about how the student would measure business performance at the company and what information/metrics they would use. Look for answers that show the candidate did their research and has a good sense of the company's goals. Also, look for signs that the candidate can adopt a business mindset and is familiar with the industry's practices and norms.
>
> 3. Ask a technical question that assesses the student's skills as they relate to the job requirements and/or required qualifications.
>
> 4. Ask another technical question that assesses the student's software knowledge as it relates to the job requirements and/or required qualifications.
>
> 5. Ask a situational question that assesses the student's ability to work in a team and/or handle difficult situations. Make the question tailored to what would be expected in this job. Possible questions can include, but are not limited to:
>    – Tell me about a time when you think you demonstrated good data sense.
>    – Describe your most complex data project from start to finish. What were the most difficult challenges, and how did you handle them?
>    – Tell me about a time when you had to work with a difficult team member. How did you handle the situation?
>
> 6. Ask a behavioral question to screen for their soft skills. Example questions to ask include, but are not limited to:
>    – What do you think are the three best qualities that great data analysts share?
>    – How would you explain your findings and processes to an audience who might not know what a data analyst does?

**Listing 4. System Prompt for the Interview Mentor Module in ChatISA.**

# 4    ChatISA Evaluation: Use Cases of ChatISA

This section provides two examples of how our undergraduate students used ChatISA in their classes. The examples are not meant to be exhaustive but were selected to show the utility of ChatISA across our curriculum.

## 4.1    Coding Companion for Data Visualization

In the *Business Intelligence and Data Visualization* course, a required course for undergraduate ISA students, students were provided with a sample dataset from the *American Community Survey (ACS)*. The ACS is administered by the US Census Bureau and is given every year to a random sample of about 3.5 million households (about 3% of all US households). Data on a random sample of 1% of all US residents are made public. For the assignment question, the students were provided with a down sampled dataset of 505 observations and 9 variables from the ACS data that included the variables: (a) *sex*, where Female = '1' and Male = '0', and (b) *income*, which contains the combined wages and salaries for each sampled individual over the past 12 months (in $1,000s). The students were asked to use any software of their choice to plot the relationship between sex and income. Note that the unit of analysis in the dataset is a sampled individual (i.e., every row corresponds to a person). The first four rows of the dataset are provided in **Error! R eference source not found.** for the sake of completeness.

**Table 1. A Sample of the First Four Rows Shared with the Students.**

| sex | income |
|-----|--------|
| '0' | 60 |
| '1' | 0.36 |
| '1' | 87 |
| '0' | 0 |

Student Prompt: "What graph is most effective to visualize the relationship between sex (categorical variable) and income (continuous) and why is it more effective than other visualizations?"

ChatISA's Response: "We reran the student's initial prompt using the latest version of ChatISA, with the GPT-4o model (our default LLM). Figure 5 shows the entire output obtained from ChatISA. The animated output requires a non-browser-based PDF reader. Alternatively, the reader can view a recording of the interaction at https://www.loom.com/share/f7ad734229f646c6aefcd1042882fc4e?sid=5cfa2d5f-101f-40e2-848f-0330c11243ce."

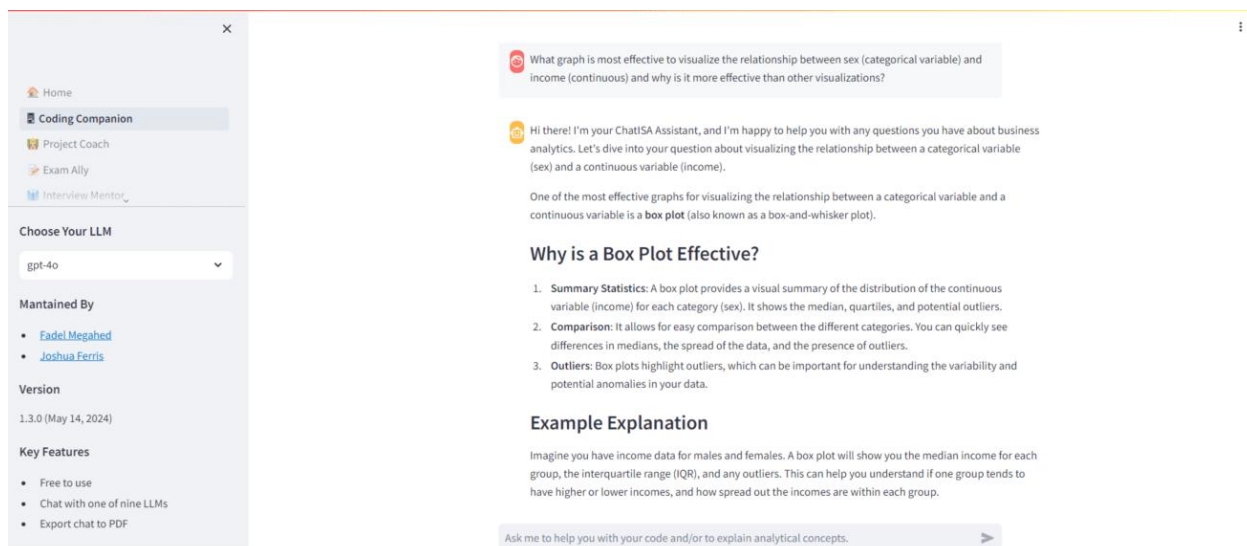

**Figure 5. ChatISA's generated response for the data visualization prompt.**

The student and the faculty agree that the generated answer is reasonable. The answer presents a box plot as its preferred solution over: (a) bar charts, which are suboptimal since they would only show the mean/median, hiding information about the distribution and outliers (still true with the introduction of error bars since they do not capture the entire distribution); (b) scatter plots since they are meant to plot the relationship between two categorical variables, and (c) histograms, which are typically used for a single continuous variable (even though we know that we can overlay two histograms by making the bins semi-transparent). Due to our prompt engineering, the answer provides examples of how to construct box plots in both R and Python, with the R code clearly matching our suggested coding instructions. Furthermore, ChatISA attempted to provide an additional explanation for why the box plot would be suitable by stating, "*Imagine you have income data for males and females. A box plot will show you the median income for each group, the interquartile range (IQR), and any outliers. This can help you understand if one group tends to have higher or lower incomes, and how spread out the incomes are within each group.*" In an attempt to capture the impact of our temperature setting (set to 0 for the coding companion) and our system prompt on the generated answer, we reran the prompt in ChatGPT (https://chat.openai.com/) using the GPT-4o model. ChatGPT's answer was mostly the same by recommending the boxplot. As expected, the ChatGPT answer did not include code (since the prompt did not ask for it) and did not provide an example explanation as above. In its comparison with other charts, ChatGPT excluded scatter plots and added Violin and Strip plots to the comparison, noting that both provide distributional information but are harder to interpret. When we ran it again in ChatGPT to see if this had to do with the stochastic nature of LLMs, the scatter plot was still excluded, and ChatGPT added a bar chart with errors (instead of the bar chart) while still having the same perceived limitations of the Violin and Strip plot.

## 4.2    Exam Ally to Prepare for the Experimental Design Final Exam

To supplement his studying for the final exam of the *Statistical Monitoring and Design of Experiments* course, an elective for the Business Analytics major, a student compiled an RMarkdown document (which was knitted/saved as a PDF) of all the course assignment questions and their solution and used them to generate multiple-choice review questions for the course. The compiled PDF, titled "AllHomeworks.pdf", can be accessed via our GitHub Repository.

**Modified Student Prompt:** "Based on the attached document, please ask me questions related to the general concepts and not the specific experiments since the final exam will involve experiments that we did not see in the assignments." Note that this prompt was modified from the student's original prompt to provide the AI with further guidance so that the questions are not specific to the datasets and experiments in the assignment.

**ChatISA's Response:** "We reran the modified prompt using the latest version of ChatISA, with the GPT-4o model (our default LLM). Figure 6 shows the entire output obtained from ChatISA. Similar to the earlier example, the animated output requires a non-browser-based PDF reader (if not available, we refer the reader to the exported PDF output from this interaction, which is made available in our GitHub Repository). Alternatively, the reader can view a recording of the interaction with `ChatISA` at https://www.loom.com/share/239950fad0e24ef1875e8d5fb35cbe60?sid=6f3679bc-5f4a-4c08-8146-b6d26cd0dc4c."

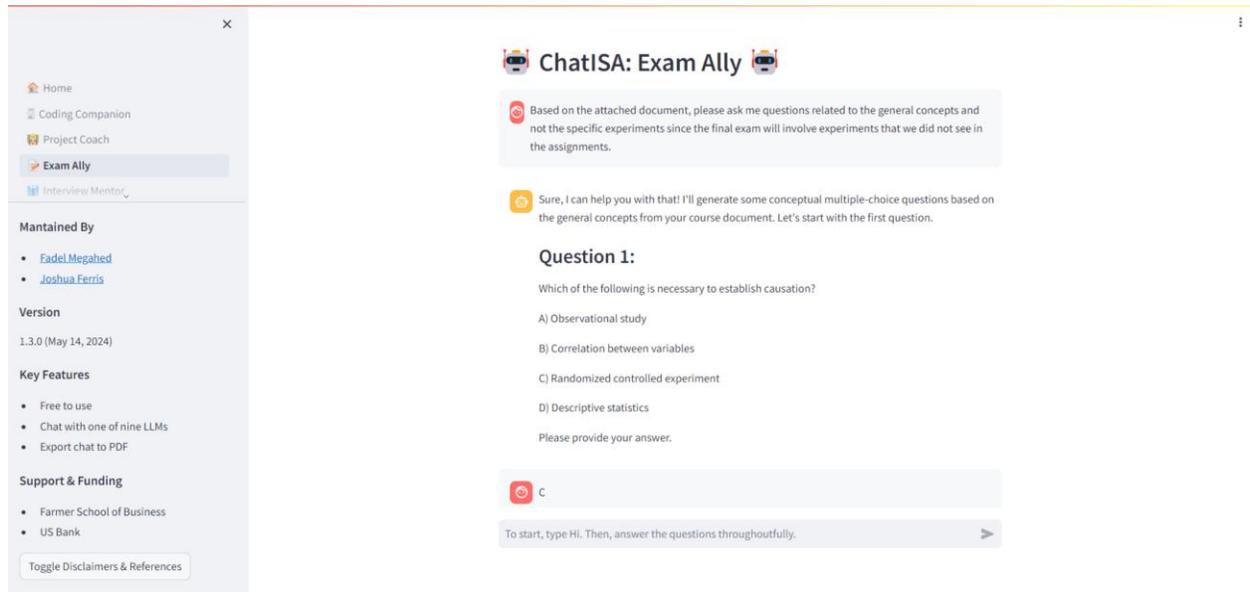

**Figure 6. ChatISA's Generated Response for the Exam Ally Prompt.**

Students generally enjoyed the interactive nature of *Exam Ally.* The questions covered topics discussed in the assignments, and the AI was able to identify the correct answer for each question. The user was presented with one question at a time and received feedback before moving to the next question. However, it is unclear whether this would significantly improve the assessments since they will still be constrained by the information uploaded by the students. In our example, one student compiled their previous assignment solutions and asked the AI to generate questions based on that document, while the other student utilized the instructor-provided study guide for her Python Programming course and found that the AI-generated questions were similar to what she encountered in her final exam. Despite the students' excitement about *Exam Ally*, we thus noted and emphasized to our students that this tool should be used only as a final stage in their exam preparation.

## 5 Discussion and Implications

Generative AI has been profoundly influenced by the release of ChatGPT on November 30, 2022, prompting extensive research into its applications and potential misuse in education, research, and professional practice (Megahed et al., 2024). In response to concerns surrounding generative AI's limitations, researchers have developed AI chatbots that incorporate grounding mechanisms to mitigate hallucinations and enhance the accuracy of responses (Megahed et al., 2024). Aligning with these efforts, this study introduces ChatISA, an in-house, multi-modal generative AI chatbot designed specifically for IS education. The primary objectives of ChatISA are twofold: first, to provide IS students with an accessible and cost-effective opportunity to engage with generative AI chatbots for educational purposes, and second, to offer IS faculty a framework for developing and integrating in-house generative AI solutions into the curriculum. This approach addresses the growing need to equip students with hands-on experience in Generative AI, which is expected to play an increasingly critical role in their academic and professional careers.

The development of ChatISA has yielded several key insights that underscore its theoretical, design, and practical contributions to IS education. These insights provide valuable guidance for future projects involving generative AI in educational contexts and highlight both the successes and challenges encountered during its implementation.

First, this study contributes to IS scholarship by designing and developing generative AI applications grounded in theoretical frameworks. Drawing upon the well-established Task-Technology Fit (TTF) theory, ChatISA was developed as a theory-driven, rigorous generative AI system aimed at enhancing students' task performance. As summarized in Table 2, we identified four primary educational tasks where IS students could derive substantial benefits from generative AI. To support these tasks, ChatISA was designed in alignment with usability guidelines (Lee & Kozar, 2012), ensuring an interface that is consistent, easy to learn, readable, interactive, flexible, and trustworthy. The system includes key features such as a streamlined user interface, custom PDF generation using Streamlit, and integration with the LangChain library to facilitate reliable data retrieval from multiple LLMs. Additionally, ChatISA is built with a high-resilience mechanism to address potential LLM outages; if one model experiences downtime, students can seamlessly switch to an alternative LLM via a dropdown menu, ensuring uninterrupted use. Preliminary observations indicate that ChatISA generates accurate and relevant outputs at no additional cost to students, fostering significant interest, frequent usage, and positive feedback. Student engagement is reflected in assignment usage patterns, informal feedback, and API usage metrics, demonstrating an increased reliance on ChatISA for academic support. Although the chatbot requires further refinement to expand task coverage, enhance usability, and improve accuracy through additional grounded data and advanced LLMs, it has already proven to be an effective AI assistant. Furthermore, it has facilitated meaningful discussions on the implications of AI in education and future professional settings, reinforcing the validity of TTF theory as a guiding framework for generative AI design in IS education.

**Table 2. A Summary of Task-Technology Fit for ChatISA.**

| Task | Technology | Task-Technology Fit |
|---|---|---|
| Coding Companion: assist students with coding-related inquiries | Streamlit (1.42.2) for the web-based graphical user interface that provides a consistent look across four tasks<br><br>The dropdown menu for choosing various large language models (LLMs) offered by OpenAI, Anthropic, Mistral, and Cohere<br><br>LangChain (0.3.19) library to have a consistent API for invoking and retrieving data from the APIs of multiple LLMs<br><br>Streamlits expander for custom export to PDF<br><br>The system prompts for asking questions and retrieving code snippets | Offer an interactive and personalized learning environment that enables students to actively learn by asking questions and tailoring explanations based on their academic backgrounds and preferred coding styles, and current understanding and learning needs<br><br>Create a collaborative coding experience by using the output from one model as input to another in an iterative manner<br><br>Adopt the dual-language approach by providing code snippets in both R and Python that accommodate diverse coding preferences within the business analytics program and ensure students gain proficiency in both languages<br><br>Set temperature to 0 for all the LLMs |
| Project Coach: steer students through project-related challenges | Streamlit(1.42.2), LLMs, LangChain, PDF<br><br>The system prompt for the specific role (Scoping Coach, Premortem Coach, Team Structuring Coach, Devils Advocate, and Reflection Coach) | Provide the specific system prompts to ensure the interaction is tailored to the specific role (Scoping Coach, Premortem Coach, Team Structuring Coach, Devil's Advocate, and Reflection Coach)<br><br>Integrate the prompts into a single tool that allows students to experiment with different coach roles using various LLMs.<br><br>Set temperature to 0 for all the LLMs |
| Exam Ally: assist students preparing for their information systems and analytics exams | Streamlit(1.42.2), LLMs, LangChain, PDF<br><br>The dropdown menu for the type of generated questions including Conceptual Multiple Choice, Conceptual Short Answer, Code Understanding, or Data Analysis | Assist with exam preparation by generating questions based on information extracted from uploaded PDFs of course materials<br><br>Enable to specify the type of generated questions<br><br>Set the temperature parameter to 0.25 to allow for some creativity in the exam questions |
| Interview Mentor: help students prepare for job interviews | Streamlit(1.42.2), LLMs, LangChain, PDF<br><br>Dropdown menus for model choice, grade and major, a PDF upload for the resume and text inputs for job title and job description | Facilitate the generation of tailored interview questions based on user-input job descriptions, interviewee info (major, grade), and an uploaded resume in PDF format<br><br>Set the temperature to 0.25 |

Second, this study also contributes to the advancement of IS programs by offering an accessible, in-house generative AI solution that can be seamlessly integrated into curricula. Unlike previous generations of students, today's learners navigate an educational landscape deeply intertwined with digital technologies. The transition toward competency-based curricula emphasizes the importance of students acquiring the ability to translate knowledge, skills, and abilities into valued organizational outcomes (Chyung et al., 2006). The integration of ChatISA into IS education fosters experiential learning, enabling students to enhance their adaptability and job performance through direct engagement with generative AI technology. Moreover, as noted by Vázquez-Cano (2021), embracing generative AI in education aligns with UNESCO's "2030 Agenda for Sustainable Development," particularly Sustainable Development Goal 4 (SDG4), which seeks to ensure inclusive and equitable quality education and promote lifelong learning opportunities. ChatISA also presents new opportunities for faculty development, allowing instructors to leverage AI-generated learning analytics to tailor instructional strategies. By analyzing student interactions and performance data, educators can identify trends, optimize pedagogical approaches, and design more personalized learning experiences (Kumar et al., 2024). Additionally, generative AI can support innovative teaching methodologies such as problem-based learning and flipped classrooms, enhancing engagement and fostering creativity in educational practices (Neumann et al., 2023). Beyond individual learning outcomes, ChatISA also redefines the role of AI in business education by shifting the student mindset from perceiving AI as a replacement for human intelligence to viewing it as a collaborator or co-pilot in problem-solving. This perspective aligns with the widely cited assertion that "AI will not replace humans, but humans with AI will replace humans without AI" (Lakhani & Ignatius, 2023). By fostering this understanding, ChatISA contributes to developing an AI-literate workforce prepared for the evolving demands of AI-integrated business environments.

Third, inclusivity and accessibility are central to ChatISA's design. The chatbot features a user-friendly interface that accommodates individuals with varying levels of technological proficiency, ensuring that all students, regardless of prior experience, can interact with the system effectively. ChatISA enhances students' autonomy and control over their learning experience by offering multiple LLM model options, allowing them to tailor their interactions to meet specific academic and personal preferences. By open-sourcing ChatISA, this study directly contributes to the IS community by demonstrating the feasibility of cost-effective, in-house generative AI solutions. Prior research has highlighted the financial and technical barriers to developing custom AI chatbots (Reddy et al., 2024). ChatISA addresses these concerns by providing a replicable, affordable model, ensuring that students who may not have access to premium chatbot services can still benefit from generative AI in their education.

Finally, while the implementation of ChatISA represents a significant step forward, several challenges remain in AI-assisted teaching and learning. One primary concern is the risk of student misuse, including potential reliance on AI for completing assignments dishonestly, thereby compromising academic integrity. To address this, clear ethical guidelines must be established, and faculty should emphasize the role of AI as a supplementary tool rather than a substitute for critical thinking. Additionally, ChatISA's effectiveness depends on institutional willingness to adapt pedagogical practices. Many faculty members and students currently engage with AI tools experimentally, yet the challenge lies in transitioning from casual AI use to systematic and intentional curricular integration (Microsoft & LinkedIn, 2024). Given that over 75% of knowledge workers now use AI at work (Microsoft & LinkedIn, 2024), IS programs must carefully consider how AI tools should be incorporated into classroom instruction. For example, we may debate whether our introductory ISA classes should fully utilize LLMs (whether customized or not) in translating the course material into code. Should we teach more content now, given that students have AI assistants? Should assignments ask students to program in multiple languages? How much should we push students? As the paradigm is changing, how should we leverage it to make our students more prepared? We do not necessarily have answers to these questions, but the ISA faculty will keep these considerations on the table. This complex task should be at the forefront of higher education discussions. Another key challenge is maintaining student agency in learning. Excessive reliance on AI-generated responses risks eroding problem-solving skills and critical engagement (Woods, 2024). The co-pilot metaphor used to describe AI-assisted learning may obscure the complex distribution of agency, leading students to over-delegate tasks to AI. Educators must balance AI integration with traditional pedagogical practices to ensure that students continue to develop independent problem-solving capabilities.

In conclusion, while ChatISA has demonstrated promising results in supporting IS education, its continued development and refinement are necessary to maximize its benefits while mitigating potential drawbacks. Addressing concerns related to ethical AI use, curricular adaptation, and student agency will be essential in ensuring that generative AI enhances, rather than disrupts, the learning process.

# 6   Limitations and Future Research

This study has several limitations that should be addressed in future research. First, as AI models used in chatbots continue to advance and the tasks undertaken by ISA students become increasingly diverse, ChatISA must be continuously updated to integrate the latest LLMs and more proprietary data and enhance its usability. Additionally, it is essential to align the system with ethical AI principles, including transparency, fairness, trust, and privacy (Chen et al., 2023; Chen et al., 2024). Second, while the primary objective of this study is to develop and propose a generative AI system, rigorous empirical research is necessary to validate its effectiveness beyond the current preliminary use case analyses. Specifically, future studies grounded in task-technology fit theory should conduct hypothesis testing and empirical validation to assess the alignment between the diverse tasks ChatISA supports and its corresponding technological features.

# 7   Conclusion

As companies and higher education integrate generative AI into their business operations and curriculum at an exponential and dramatic speed, there is an urgent need to incorporate generative AI tools in the ISA curriculum. The proliferation of generative AI technologies has contributed to the emergence of new fields and job titles, reflecting the dynamic nature of the modern workplace, although certain job roles have evolved or been replaced by the technologies. Our study develops ChatISA, an in-house generative AI chatbot, and provides a detailed description of how to build the system and presents its use cases with its four modules. We demonstrate the application of generative AI in enhancing the learning experiences of ISA students and explore its implications for updating IS curricula to effectively integrate this technology. While challenges such as ethical guidelines and the need for continuous updates remain, the positive impact on student engagement and learning is evident. By making all our code publicly available on GitHub, we hope to encourage other institutions to adopt and customize ChatISA, fostering a collaborative effort to integrate AI into educational curricula and improve learning outcomes across diverse contexts.

## Data Available Statement

To encourage extensions to our app, we made the code used for the app publicly available on GitHub with an MIT license. Our GitHub repository is publicly available at https://github.com/fmegahed/chatisa, and the chatbot is available at https://chatisa.fsb.miamioh.edu/.

## Declaration of AI

*During the preparation of this work, the authors used GitHub CoPilot, ChatGPT, and Claude to assist with syntax and debugging the programming code. They also edited the text using Grammarly and ChatGPT.*

## References

AACSB. (2025). GenAI Adoption in Business Schools: Deans and Faculty Respond. https://www.aacsb.edu/insights/reports/2025/genai-adoption-in-business-schools-deans-and-faculty-respond

Abbas, R., Papagiannidis, S., Mäntymäki, M., Hong, Y., & Lakshmi, V. (2024). Special Issue Call for Papers - Digital Transformation of Higher Education: Pedagogical Innovation Toward Education 4.0? *Communications of the Association for Information Systems, 55*, pp. 1-8.

Al-Amin, M., Ali, M. S., Salam, A., Khan, A., Ali, A., Ullah, A., Alam, M. N., & Chowdhury, S. K. (2024). History of generative Artificial Intelligence (AI) chatbots: past, present, and future development. *arXiv preprint arXiv:2402.05122.*

Alavi, M., Leidner, D., & Mousavi, R. (2024). Knowledge management perspective of generative artificial intelligence (GenAI). *Alavi, Maryam*, pp. 1-12.

Albeedan, M., Kolivand, H., & Hammady, R. (2023). Evaluating the Use of Mixed Reality in CSI Training through the Integration of the Task-Technology Fit and Technology Acceptance Model. *IEEE Access, 11*, pp. 114732--114752.

Anderson, C., Carvalho, A., Kaul, M., & Merhout, J. W. (2023). Blockchain innovation for consent self-management in health information exchanges. *Decision Support Systems, 174*, p. 114021.

Anthropic. (2024). Prompt generator. https://docs.anthropic.com/en/docs/prompt-generator

Anthropic. (2025). Claude 3.7 Sonnet and Claude Code. https://www.anthropic.com/news/claude-3-7-sonnet

Artifex Software Inc. (2024). PyMuPDF4LLM Documentation. Artifex Software Inc.

Bisgambiglia, P.-A., Nivet, M.-L., & Vittori, E. (2024, Feb). *How to shape Computer Science Education in the AI Era? Bridging Technology, Humanities, and Inspiring the Desire to Learn.* https://undonecs.sciencesconf.org/data/pages/07_02_Marie_Laure_Nivet.pdf

Chen, H., & Anyanwu, C. C. (2025). AI in education: Evaluating the impact of moodle AI-powered chatbots and metacognitive teaching approaches on academic performance of higher Institution Business Education students. *Education and Information Technologies*, pp. 1-16.

Chen, J.-S., Tran-Thien-Y, L., & Florence, D. (2021). Usability and responsiveness of artificial intelligence chatbot on online customer experience in e-retailing. *International Journal of Retail & Distribution Management, 49*(11), pp. 1512-1531.

Chen, R., Feng, J., de Matos, M. G., Hsu, C., & Rao, H. R. (2024). Artificial Intelligence-Information Assurance Nexus: Shaping the Future of Information Systems Security, Privacy, and Quality. https://misq.umn.edu/call-for-papers-ai-ia

Chen, Y., Jensen, S., Albert, L. J., Gupta, S., & Lee, T. (2023). Artificial intelligence (AI) student assistants in the classroom: Designing chatbots to support student success. *Information Systems Frontiers, 25*(1), pp. 161-182.

Chukwuere, J. E., & Handoko, B. L. (2024). The future of generative AI chatbots in higher education. *Journal of Emerging Technologies, 4*(1), pp. 36-44.

Chyung, S. Y., Stepich, D., & Cox, D. (2006). Building a competency-based curriculum architecture to educate 21st-century business practitioners. *Journal of Education for Business, 81*(6), pp. 307-314.

Claude. (2024). Claude Upgrade. https://claude.ai/upgrade

Coffey, L. (2024). Majority of Grads Wish They'd Been Taught AI in College. *Inside Higher Ed.* https://www.insidehighered.com/news/tech-innovation/artificial-intelligence/2024/07/23/new-report-finds-recent-grads-want-ai-be

Echterhoff, J., Liu, Y., Alessa, A., McAuley, J., & He, Z. (2024). Cognitive bias in decision-making with LLMs. *arXiv preprint arXiv:2403.00811.*

Flaherty, C. (2025). The Digital Divide: Student Generative AI Access. *Inside Higher Ed.*

Fortune Business Insights. (2025). Generative AI Market Size, Share & Industry Analysis. https://www.fortunebusinessinsights.com/generative-ai-market-107837.

Garn, D. (2025). How to Use Perplexity AI: Tutorial, Pros and Cons. *TechTarget.*

George, B., & Wooden, O. (2023). Managing the strategic transformation of higher education through artificial intelligence. *Administrative Sciences, 13*(9), p. 196.

Goodhue, D. L., & Thompson, R. L. (1995). Task-technology fit and individual performance. *MIS quarterly*, pp. 213-236.

Google. (2024). Interview Warmup. https://grow.google/certificates/interview-warmup/

Gottschalk, F., & Weise, C. (2023). Digital equity and inclusion in education: An overview of practice and policy in OECD countries. Organisation for Economic Co-operation and Development. DIRECTORATE FOR EDUCATION AND SKILLS. OECD Education Working Paper No. 299. https://one.oecd.org/document/EDU/WKP(2023)14/en/pdf

Grassini, S. (2023). Shaping the future of education: Exploring the potential and consequences of AI and ChatGPT in educational settings. *Education sciences, 13*(7), p. 692.

Gupta, R., Nair, K., Mishra, M., Ibrahim, B., & Bhardwaj, S. (2024). Adoption and impacts of generative artificial intelligence: Theoretical underpinnings and research agenda. *International Journal of Information Management Data Insights, 4*(1), p. 100232.

Hirabayashi, S., Jain, R., Jurković, N., & Wu, G. (2024). Harvard undergraduate survey on generative AI. *arXiv preprint arXiv:2406.00833*.

Huang, H., Li, Y., Jiang, B., Liu, L., Jiang, B., Sun, R., Liu, Z., & Liang, S. (2025, April). On-Premises LLM Deployment Demands a Middle Path: Preserving Privacy Without Sacrificing Model Confidentiality. *ICLR 2025 Workshop on Building Trust in Language Models and Applications*.

Huy, L. V., Nguyen, H. T., Vo-Thanh, T., Thinh, N. H., Thi Thu Dung, T., & others. (2024). Generative AI, why, how, and outcomes: A user adoption study. *AIS Transactions on Human-Computer Interaction, 16*(1), pp. 1-27.

IBM Research. (2023). What is Generative AI? https://research.ibm.com/blog/what-is-generative-AI

Ilieva, G., Yankova, T., Klisarova-Belcheva, S., Dimitrov, A., Bratkov, M., & Angelov, D. (2023). Effects of generative chatbots in higher education. *Information, 14*(9), p. 492.

International Organization for Standardization. (2018). Ergonomics of Human System Interaction: Usability, Definitions and Concepts. https://www.iso.org/standard/63500.html

Jeon, J. (2024). Exploring AI chatbot affordances in the EFL classroom: Young learners' experiences and perspectives. *Computer Assisted Language Learning, 37*(1-2), pp. 1-26.

Kaufman, M. (2025). It's America's Fastest-Growing Job – Thanks to ChatGPT. *CNN Business.*

Kelly, R. (2024, August 28). Survey: 86% of Students Already Use AI in Their Studies. *Campus Technology.*

Kim, J. S., Kim, M., & Baek, T. H. (2025). Enhancing user experience with a generative AI chatbot. *International Journal of Human–Computer Interaction, 41*(1), pp. 651-663.

Kirrane, S. (2025). *The power of many: The benefits of multiple LLMs.* Narus AI. https://www.narus.ai/blog/the-power-of-many-the-benefits-of-multiple-llms

Kumar, S., Rao, P., Singhania, S., Verma, S., & Kheterpal, M. (2024). Will artificial intelligence drive the advancements in higher education? A tri-phased exploration. *Technological Forecasting and Social Change, 201*, p. 123258.

Lakhani, K., & Ignatius, A. (2023). AI won't replace humans–but humans with AI will replace humans without AI. *Harvard business review, 4.*

Lee, Y., & Kozar, K. A. (2012). Understanding of website usability: Specifying and measuring constructs and their relationships. *Decision support systems, 52*(2), pp. 450-463.

Lin, Z., Guan, S., Zhang, W., Zhang, H., Li, Y., & Zhang, H. (2024). Towards trustworthy LLMs: a review on debiasing and dehallucinating in large language models. *Artificial Intelligence Review, 57*(9), p. 243.

Maudsley, D. B. (1979). *A theory of meta-learning and principles of facilitation: An organismic perspective.* University of Toronto, 27 King's College Cir, Toronto, ON M5S 1A1, Canada.

Megahed, F. M., Chen, Y.-J., Ferris, J. A., Knoth, S., & Jones-Farmer, L. A. (2024). How generative AI models such as ChatGPT can be (mis)used in SPC practice, education, and research? An exploratory study. *Quality Engineering, 36*(2), pp. 287-315.

Megahed, F. M., Chen, Y.-J., Zwetsloot, I., Knoth, S., Montgomery, D. C., & Jones-Farmer, L. A. (2024). Introducing ChatSQC: Enhancing Statistical Quality Control with Augmented AI. *Journal of Quality Technology, 56*(5), 474-497.

Microsoft, & LinkedIn. (2024). AI at Work Is Here. Now Comes the Hard Part. 2024 Work Trend Index Annual Report from Microsoft and LinkedIn. https://assets-c4akfrf5b4d3f4b7.z01.azurefd.net/assets/2024/05/2024_Work_Trend_Index_Annual_Report_63d45200a4ad.pdf

Mollick, E., & Mollick, L. (2023). Assigning AI: Seven approaches for students, with prompts. *arXiv preprint arXiv: 2306.10052.*

Murray, S. (2025). Business Schools Ease Their Resistance to AI. *Financial Times.*

Neumann, M., Rauschenberger, M., & Schön, E.-M. (2023). "We need to talk about ChatGPT": The future of AI and higher education. *2023 IEEE/ACM 5th International Workshop on Software Engineering Education for the Next Generation (SEENG)* (pp. 29-32). IEEE.

Nguyen, Q. N., Sidorova, A., & Torres, R. (2022). User interactions with chatbot interfaces vs. Menu-based interfaces: An empirical study. *Computers in Human Behavior, 128*, p. 107093.

Nielsen, J. (2023). ChatGPT Lifts Business Professionals' Productivity and Improves Work Quality. *Nielsen Norman Group*.

OhioLink. (2022). Introduction to Affordable Learning Ohio. OhioLink: Ohio A Division of the Ohio Department of Higher Education. https://www.ohiolink.edu/content/affordablelearning

OpenAI. (2022). Introducing ChatGPT. https://openai.com/index/chatgpt

Park, K. (2023). Samsung Bans Use of Generative AI Tools Like ChatGPT After April Internal Data Leak. *TechCrunch*.

Peffers, K., Tuunanen, T., Rothenberger, M. A., & Chatterjee, S. (2007). A design science research methodology for information systems research. *Journal of management information systems, 24*(3), pp. 45-77.

Reddy, N. S., Chaitanya, N. P., Varshitha, P. S., Varma, R. C., Reddy, P. K., & Dandu, J. (2024). Intelligent Chatbot For Educational Institutions. *2024 7th International Conference on Circuit Power and Computing Technologies (ICCPCT). 1*, pp. 1337-1343. IEEE.

Riapina, N. (2024). Teaching AI-enabled business communication in higher education: A practical framework. *Business and Professional Communication Quarterly, 87*(3), pp. 511-521.

Rienties, B., Domingue, J., Duttaroy, S., Herodotou, C., Tessarolo, F., & Whitelock, D. (2024). What distance learning students want from an AI Digital Assistant. *Distance Education*, pp. 1-17.

Roth, T., Stohr, A., Amend, J., Fridgen, G., & Rieger, A. (2023). Blockchain as a driving force for federalism: A theory of cross-organizational task-technology fit. *International Journal of Information Management, 68*, p. 102476.

Seifi, N., & Chugh, M. (2025). Multi-LLM Routing Strategies for Generative AI Applications on AWS. https://aws.amazon.com/blogs/machine-learning/multi-llm-routing-strategies-for-generative-ai-applications-on-aws/

Starr, D., Hayes, J., & Gao, N. (2022). The Digital Divide in Education. Public Policy Institue of California. Fact Sheet - June 2022. https://www.ppic.org/wp-content/uploads/the-digital-divide-in-education.pdf

Under, C. D. (2025). *GPT-4.1 vs Claude 3.7 vs Gemini 2.5 Pro vs Grok 3: The four horsemen of the AI revolution.* Medium. https://medium.com/@cognidownunder/gpt-4-1-vs-claude-3-7-vs-gemini-2-5-pro-vs-grok-3-the-four-horsemen-of-the-ai-revolution-4fbcef192b11

Vázquez-Cano, E. (2021). Artificial intelligence and education: A pedagogical challenge for the 21st century. *Educational Process: International Journal, 10*(3), pp. 7-12.

Wang, S. L., & Lin, H. I. (2019). Integrating TTF and IDT to evaluate user intention of big data analytics in mobile cloud healthcare system. *Behaviour & Information Technology, 38*(9), pp. 974-985.

Waters, R. (2025). The Diverging Future of AI. *Financial Times*.

Woods, K. (2024). If AI is our co-pilot, who is the captain? *AI & SOCIETY*, pp. 1-2.

Wu, R., & Yu, Z. (2024). Do AI chatbots improve students learning outcomes? Evidence from a meta-analysis. *British Journal of Educational Technology, 55*(1), pp. 10-33.

Yang, S., & Evans, C. (2019). Opportunities and challenges in using AI chatbots in higher education. *Proceedings of the 2019 3rd International Conference on Education and E-Learning*, (pp. 79-83).

Yeon, J., Jung, Y., Baek, Y., Lee, D., Shin, J., & Chung, W. Y. (2024). User Preferences on a Generative AI User Interface Through a Choice Experiment. *International Journal of Human–Computer Interaction*, pp. 1-12.

Zhang, D., Sanyal, P., Nah, F. F.-H., & Mukkamala, R. (2024). Generative AI: Transforming Human, Business, and Organizational Decision-Making. https://www.sciencedirect.com/special-issue/299249/generative-ai-transforming-human-business-and-organizational-decision-making